\documentstyle[12pt]{article}
\newcommand{\beq}{\begin{equation}}
\newcommand{\eeq}{\end{equation}}
\newcommand{\bea}{\begin{eqnarray}}
\newcommand{\eea}{\end{eqnarray}}

\renewcommand{\thefootnote}{\fnsymbol{footnote}}
\begin{document}
\topmargin=0.0truecm
\oddsidemargin=-0.8truecm
\evensidemargin=-0.8truecm
\setcounter{page}{1}
\vspace*{3.0cm}
\begin{center}
SOLAR NEUTRINOS AND LEPTON MIXING\\

\vspace*{0.1cm}
A. Yu. Smirnov\\
\vspace*{-0.1truecm}  \mbox{}
\noindent
International Centre for Theoretical Physics, I-34100, Trieste, Italy
and\\
\vspace*{-0.2truecm}  \mbox{}
Institute for Nuclear Research, RAS,
107 370 Moscow, Russia\\
\vspace*{-0.3truecm}
\end{center}
\vspace*{6.5cm}
\begin{center}
ABSTRACT
\end{center}
With latest experimental data the solar neutrino problem enters
new phase when crucial aspects of the problem
can be formulated in an essentially (solar) model independent way.
Original neutrino fluxes can be considered as free
parameters to be found from the solar neutrino experiments.
Resonance flavour conversion gives the best
fit of all experimental results. Already
existing data allow one to constraint  both the neutrino
parameters and the original neutrino fluxes. The reconciliation
of the  solution of the solar neutrino problem
with other neutrino mass hints  (atmospheric neutrino problem,
hot dark matter etc.) may require the existence of
new very light singlet fermion. Supersymmetry can provide
a framework within which the desired properties of such a light
fermion follow naturally. The existence of the fermion
can be related to axion physics, mechanism of
$\mu$-term generation etc..

\renewcommand{\thefootnote}{\arabic{footnote}}
\setcounter{footnote}{0}

\newpage
\noindent
{\large 1. Introduction.}\\

There are  three phases in understanding of the
solar neutrino problem:

\noindent
1. {\it Theory without experiment}. The problem was predicted by
B. Pontecorvo: Even before the first Homestake results
he  suggested that  neutrino oscillations can influence
the solar neutrino fluxes, diminishing the detected signals.\\
2. {\it Theory and Experiment}. During more than 20 years
we had one experiment and one model. Namely, Davis's experiment
and Bahcall's predictions of neutrino fluxes in the
Standard Solar Model (SSM).
The problem was formulated as the smallness of the
Homestake signal in comparison with Bahcall's prediction.\\
3. {\it Experiment without theory}. Results (now rather precise)
from Homestake$^{1)}$,
Kamiokande$^{2)}$, SAGE$^{3)}$ and GALLEX$^{4)}$
experiments as well as  calibration
of  Kamiokande and  recent GALLEX experiment with $^{51}Cr$ source allow
one to formulate the problem in almost solar model independent way.

There is some hope that forthcoming experiments
SuperKamiokande and SNO will resolve the problem
(at least establish, finally, whether the astrophysics or
neutrino properties are responsible for the observable deficit)
without referring to the predictions of the specific solar models.

In this paper we summarize  essential points of this third phase
and consider some  implications of the data to the lepton mixing.\\

\noindent
{\large 2. Solar Neutrino Problem without Solar Neutrino
Model.}\\

In spite of serious  progress in the solar modeling and  very good
agreement of SSM and helioseismological data, the predicted solar
neutrino fluxes  still have rather large uncertainties.  Mainly,
they are related to the nuclear cross-sections (first of all, for  the
reaction  $p + ^7 Be \rightarrow ^8 B + \gamma$)
and probably to some plasma effects
which have not yet been properly taken into account.
These uncertainties will hardly be fixed before new experiments on solar
neutrinos  start to operate.
In this connection  it is instructive to  use as much as
possible model independent approach to the problem, and
try to resolve it using solar neutrino data only.
Main points of the approach are the following$^{5-18)}$.

1. Only general notion is used about the
solar neutrinos: the composition, the energy
spectra of
components, but not the absolute values of fluxes. These  absolute
values are considered as {\it free parameters to be found from the
solar neutrino experiments}.  In particular, the boron neutrino flux can be
written as
\beq
\Phi_B = f_B \cdot \Phi_B^{SSM},
\eeq
where
$f_B$ is free parameter, and $\Phi_B^{SSM}$ is the flux in the
reference SSM$^{19)}$.
Similarly, parameters $f_i$ $(i = Be, pp, NO)$ for other important
fluxes can be introduced.\\
2. One  confronts the data from different experiments immediately.\\
3. The normalization of the
solar neutrino flux is used which follows from
the solar luminosity at the condition of thermal
equilibrium of the Sun.\\

In fact, present experimental situation makes the  analysis of data
to be very simple. There are two key point in this analysis.

 {\it Kamiokande versus  Homestake}$^{5 - 16)}$. Boron neutrino flux
measured by Kamiokande gives the contribution to  Ar-production
rate $Q_{Ar}^B = 3.00 \pm 0.45$ SNU \footnote{In this estimation
it was
suggested that the contribution to the Kamiokande follows from the
electron neutrinos only.} which exceeds the total signal observed
by Homestake: $Q_{Ar}^B = 2.55 \pm 0.25$ SNU.
This means that the contributions  of all
other fluxes to $Q_{Ar}$, and in particular, of the
Beryllium neutrinos should be strongly suppressed.

{\it Gallium Experiment Results versus Solar Luminosity}$^{7,20,17)}$.
The luminosity of the Sun
allows one to estimate the
pp- neutrino flux and consequently its contribution to
Ge-production rate: $Q_{Ge}^{pp} \approx 71$ SNU. This value plus small
($\sim 5$ SNU) contribution of boron neutrinos coincide within 1$\sigma$
with total signal observed by GALLEX and SAGE. Consequently, gallium
results can be reproduced if the beryllium neutrino flux
as well as all other fluxes
of the intermediate energies are strongly suppressed.

Thus both these points indicate on strong suppression of the $^7$Be-
neutrino line. Statistical analysis gives
$f_{Be} < 0.4$ (2$\sigma$) (for more detail
see$^{13,14,15)}$).\\

It follows from the above consideration
that the data fix uniquely
values of fluxes which give the {\it best fit}$^{17)}$:

\noindent
1. Boron neutrino flux should be $\approx 0.4\Phi_{B}^{SSM}$.\\
2. Beryllium neutrino flux as well as other fluxes of the intermediate
energies  (pep, N, O) give negligible contributions to the signals.\\
3. There is little or no suppression of the pp-flux.\\
Moreover, to reproduce  central values of signals one should suggest
that there is an additional flux which contributes to the Kamiokande
signal,
$\Delta \Phi_B \approx 0.09 \Phi_B^{SSM}$, but does not contribute
to the Ar-production rate.
Any deviation from this picture gives worser fit.
Thus the energy dependence of the suppression factor $P(E)$
can be represented as
\begin{equation}
P(E) \sim \left\{
\begin{array}{ll}
0.9 - 1 & E < 0.5 \ \  {\rm MeV} \\
\sim 0       & E \sim 0.7 - 1.5 \ \  {\rm MeV}\\
0.4 - 1 & E > 7 \ \  {\rm MeV} \ .
\end{array}
\right.
\end{equation}
Large uncertainty of the suppression in high
energy region is related to
the uncertainty in the original boron neutrino flux.
Kamiokande admits a mild distortion
of the recoil electron spectrum.

Evidently the astrophysics  can not reproduce such a
picture$^{5,6,8 - 15)}$.
Typically one gets more strong suppression of the boron neutrino flux
than the beryllium neutrino flux. \\

\newpage

\noindent
{\large 3. Neutrino parameters and  neutrino fluxes}\\

There are several recent studies
of the particle physics solutions of the
solar neutrino problem for unfixed values of original
fluxes$^{16,17,18,21,22)}$.

{\it Long length vacuum oscillations} can reasonably reproduce the
desired suppression factor. For
$\Delta m^2 > 7 \cdot 10^{-11}$ eV$^2$
the pp-neutrino flux is in the
region of averaged oscillations, where
$P = 1 - 0.5 \sin^2 2\theta$, the Beryllium neutrinos are in the
fastly oscillating region of the $P(E)$ (so that one expects an appreciable
time variations of the Be-neutrino flux due to annual change of distance
between the Sun and the Earth). Boron neutrinos are in the first
(high energy) minimum of $P(E)$. This allows one to reach the
inequality $P_{pp} > P_B > B_{Be}$ implied by (2).
However,  there is an obvious relation
between maximal suppression of the Be-line and the suppression of
pp-neutrinos:
$P_{Be, min} = 2 P_{pp} -1$,
and due to this the best fit configuration (2)
can not be realized.

With diminishing $f_B$
the  needed  suppression of B-neutrino flux due to the
oscillations becomes weaker. Therefore
for fixed values of $\Delta m^2$
the allowed regions of parameters
shift to smaller $\sin^2 2\theta$ $^{21,22)}$.
In particular, for
$f_B = 0.7$,
the region is  at
$\sin^2 2\theta < 0.7$
thus satisfying the potential bound from SN87A$^{23)}$.
For  $f_B \sim  0.4$ the mixing can be as small as
$\sin^2 2\theta < 0.5 - 0.6$. Moreover, for
$f_B = 0.5$ the allowed region appears at
$\Delta m^2 \sim 5 \cdot 10^{-12}$ eV$^2$
which corresponds to a position
of the Be-neutrino line in the first high energy minimum of $P$.
Depending on neutrino parameters and
$f_B$, $f_{Be}$ ...  one can get a variety
of distortions of the boron neutrino energy
spectrum.

Being excluded at $f_B = 1$, the oscillations into sterile neutrino
are allowed for $f_B < 0.7$$^{22)}$.\\

{\it Resonance flavour conversion}
can precisely reproduce the desired energy dependence of the
suppression factor (2). In the region of small mixing angles:
\begin{equation}
P_{pp} \sim 1, \ \ \  P_{Be} \sim 0, \ \ \
P_B \sim exp(-E_{na}/E),
\end{equation}
where $E_{na} \equiv \Delta m^2 l_n \sin^2 2\theta$
An additional contribution to Kamiokande $\Delta f \approx 0.09$
follows from  scattering of the converted $\nu_{\mu}$ ($\nu_{\tau}$)
on electrons due to the neutral currents. As the result:
$R_{\nu e} \sim f_B [P_B + 1/6(1 - P_B)]$.
With diminishing $f_B$ the desired suppression due
to conversion  relaxes,  and
therefore $\sin^2 2\theta$ decreases according to (3)$^{17,18)}$.
At  $\Delta m^2 =6  \cdot 10^{-6}$ eV$^2$
the best fit of the data for flavour mixing
corresponds to$^{17)}$
\begin{equation}
\begin{array}{lllllll}
f_B            & 0.4 & 0.5 & 0.75 & 1.0 & 1.5 & 2.0 \\
\sin^2 2\theta & 1.0 \cdot 10^{-3}& 1.8 \cdot 10^{-3} &
4.3 \cdot 10^{-3} & 6.2 \cdot 10^{-3} & 9 \cdot 10^{-3}
& 10^{-2}
\end{array}
\end{equation}
(For $f_B \sim 0.38$ the best fit is at
$\Delta m^2 =4  \cdot 10^{-6}$ eV$^2$). The decrease of $f_{Be}$
gives an additional shift of the allowed region
to  smaller values of $\sin^2 2\theta$.
A consistent description
of the data has been found for$^{17)}$
$$
f_B \sim 0.4 - 2.0 .
$$
(see (4) and table in (5)). For unfixed values of the original fluxes,
$f_B$, $f_{Be}$ ...,
the allowed region of neutrino parameters  is controlled
immediately by Gallium data and by the
``double ratio"$^{17)}$. Namely,
the mass squared difference
\begin{equation}
\Delta m^2 = (6 \pm 2) \cdot 10^{-6} {\rm eV^2}, \ \ \
\end{equation}
is restricted essentially by
results from Gallium experiments which imply
that the adiabatic edge of the suppression  pit is
in between the end point of the pp-neutrino spectrum and the Be-line.
This bound does not depend on mixing angle in a wide region of $\theta$.
(For sterile neutrinos the bound is approximately the same).
For fixed $\Delta m^2$ the  mixing $\sin^2 2\theta$ is determined by
the ``double ratio''
$$
R_{H/K} \equiv {R_{Ar} \over R_{\nu_e}}\;,
$$
where
$R_{Ar} \equiv Q^{obs}_{Ar}/Q^{SSM}_{Ar}$ and
$R_{\nu_e} \equiv \Phi^{obs}_{B}/\Phi^{SSM}_{B}$ are the suppressions of
signals in $Cl$--$Ar$ and Kamiokande experiments, respectively. Here
$Q^{SSM}_{Ar}$, $\Phi^{SSM}_{B}$
are the predictions in the reference model$^{19)}$
and $Q^{obs}_{Ar}$, $\Phi^{obs}_{B}$ are the observable signals.
The ratio $R_{H/K}$ depends very weakly on the  solar model.
It has however different behaviour for the conversion
into active and into sterile neutrinos.
For $\Delta m^2 = 6\cdot 10^{-6}$ eV$^2$ we get
\begin{equation}
\begin{array}{lllll}
\sin^2 2\theta & 2\cdot 10^{-3}& 5\cdot 10^{-3} & 10^{-2}
& 2\cdot 10^{-2} \\
R_{H/K}^a      & 0.75 & 0.64 & 0.56 & 0.21 \\
R_{H/K}^s      & 0.77 & 0.74 & 0.72 & 0.69    \ \ .
\end{array}
\end{equation}
The experimental value is $R_{H/K} = 0.65 \pm 0.11$.
In the case of active neutrinos
$R_{H/K}$ drops quickly when $\sin^2 2\theta$ becomes
larger than $10^{-2}$. The reason is that   the Kamiokande signal
is dominated by NC scattering of
$\nu_{\mu}$ and $\nu_{\tau}$ and
$R_{\nu e} \to 1/6$, whereas $R_{Ar}$ is strongly suppressed.
Central value of
$R_{H/K}$ can be achieved  at
$\sin^2 2\theta = 5 \cdot 10^{-6}$ and $f_B \approx 1.1$.

In the case of conversion into sterile neutrinos
there is no NC effect for $\nu_s$ and the suppression of both
Homestake and Kamiokande signals strengthen with
$\theta$ increase  simultaneously.
As the result one has weak dependence of $R_{H/K}^s$ on mixing
angle.
However,  for large  $\sin^22\theta_{es}$ the original flux of
Boron neutrinos
should be large (to  compensate for a strong suppression effect).  If we
restrict $\Phi_B \leq  1.5 \Phi^{SSM}_B$,
then the bound on the mixing angle becomes:
$\sin^2 2\theta_{es} < 1.5\cdot 10^{-2}$.

For  very small mixing solution:
$f_B \sim 0.5$, $\sin^2 2\theta_{es} \sim 10^{-3}$,   all
the effects of
conversion in the high energy part of the boron neutrino spectrum
($E > 5 - 6$ MeV)  become very weak. In particular,
the distortion of the energy spectrum disappears, and the
ratio $(CC/NC)^{exp}/(CC/NC)^{th}$ approaches 1.
Thus studying just this
part of spectrum it will
be difficult to identify the solution (e.g., to distinguish
the conversion and the astrophysical effects).

Recent calculations in SSM
with diffusion of heavy elements give larger boron neutrino
flux$^{24)}$, so that even
with $25 \%$ decrease of nuclear cross-section  and
$2\%$ decrease of central temperature of the Sun one still needs an
appreciable conversion effect. This gives a hope that
the problem can be resolved by SuperKamiokande/SNO experiments.

With increase of $f_B$ the fit of the data in the large mixing
domain becomes better$^{17)}$
. Here the Kamiokande signal can be explained
essentially by NC effect and the mixing can be relatively small.
Be- neutrinos are sufficiently suppressed and
suppression of the pp-neutrinos is rather weak. For $f_B = 2 $ the values
$\sin^2 2\theta = 0.2 - 0.3$ become allowed. The corresponding
mass squared difference is
$\Delta m^2 = 6  \cdot 10^{-6} - 10^{-4}$ eV$^2$.\\

\noindent
{\large 4.
Lepton mixing: pattern, implications}\\

Let us consider possible implications of the solar neutrino
data to the
lepton mixing. The scale of  masses
\begin{equation}
m_2 = (2 - 3)\cdot 10^{-3} {\rm eV}
\end{equation}
needed for solar neutrinos can be
obtained by the see-saw
mechanism with the mass of the RH neutrino component
in the intermediate range: $M \sim 10^{11}$ GeV.
The common observation  is that $M_R$ can be related to
the scale of the Peccei-Quinn symmetry
breaking  or to SUSY breaking in the hidden sector etc..
The desired mixing is consistent with the following
relation
\begin{equation}
\theta_{e\mu} = \sqrt{\frac{m_e}{m_{\mu}}} - e^{i \phi} \theta_{\nu},
\end{equation}
where $\theta_{\nu}$ comes from diagonalization of  neutrino mass
matrix. The relation (8) may follow from Fritzsch ansatz
in the context of the see-saw mechanism. There are however two cautions.
In many models the angle $\theta_{\nu}$ is very small,
and from (8) one finds
$
\sin^2 2\theta_{e\mu} \approx 4 (m_e/m_{\mu}) \approx 2 \cdot 10^{-2}
$
which is too large (see table in (6)). For very small mixing solution,
$
\sin^2 2\theta_{e\mu} \sim  10^{-3},
$
one needs  strong cancellation of contributions in (8).\\

Is the solution of the solar neutrino problem  compatible with
explanations of other neutrino anomalies like
deficit of the atmospheric $\nu_{\mu}$- flux,
possible signal of the $\bar{\nu}_{\mu} - \bar{\nu}_{\tau}$ oscillations,
existence of the hot component of dark matter?
Here the key words are the ``pattern" and the ``scenarios" of
neutrino masses and mixing.
Let us outline two possibilities.\\

\noindent
{\it 1. Standard scenario of neutrino masses and mixing.}

\noindent
(i). Neutrino masses are generated by the see-saw
mechanism with
masses of the RH components $M_R = 10^{11} - 10^{12}$ GeV.
This scale can originate from  Grand Unification scale,
$M_{GU}$, and the Planck scale, $M_{Pl}$,
as $M_R \sim M_{GU}^2/M_{Pl}$.\\
(ii) Second mass, $m_2$,  is in the range (7), so that the
resonance flavour conversion $\nu_e \to \nu_{\mu}$
solves the solar neutrino problem.\\
(iii) The third neutrino (at $m^D \sim 50$ GeV and $M = 10^{12}$ GeV)
has the mass about 5 eV.
It composes the desired hot component of dark
matter.\\
(iv) The decays of the RH neutrinos with mass $10^{12}$ GeV can
produce the lepton asymmetry of the Universe
which can be  transformed in to the
baryon asymmetry during the electroweak phase transition$^{25)}$
.\\
(v) Large Yukawa coupling of neutrino from  the third generation,
e.g. $Y_{\nu} \sim Y_{top}$,
give appreciable renormalization effects in the region of momenta
$M_R - M_{GU}$. The $b - \tau$ mass ratio increases by
$(10 - 15) \%$
in the SUSY. In turn this disfavours the $b - \tau$ mass unification
for low values of $\tan \beta$ $^{26, 27)}$
.\\
(vi) Simplest schemes with  quark - lepton symmetry
lead to  mixing angle for the $e$ and $\tau$ generations:
$\theta_{e \tau} \sim (0.3 - 3) V_{ub}$ which
is close the bound from the nucleosynthesis of heavy elements
(r-processes) in the inner part of the supernova:
$\sin^2 2\theta_{e \tau} < 10^{-5}$  ($m_3 > 2$ eV)$^{28)}$.\\
(vii) For $\mu -\tau$  mixings one expects
$^{29)}$
$
\theta_{\mu \tau} \sim k V_{cb} \eta,
$
where $k = 1/3 - 3$ and $\eta \sim 1$ is the
renormalization factor.
For $m_3 > 3$ eV some part of expected region of mixing angles
is already excluded by FNAL 531.
Large part of the region can be studied by CHORUS and NOMAD. The rest
(especially $m_3 < 2$ eV) will be covered by E 803.\\
(viii) The depth of $\bar{\nu}_{\mu} - \bar{\nu}_e$
oscillations with $\Delta m^2 \approx m_3$  turns out to be
$4|U_{3\mu}|^2 |U_{3e}|^2 \approx
4|\theta _{e \tau}|^2 |\theta_{\mu \tau}|^2$. The existing
experimental data give the bound on this depth:  $ < 10^{-3}$ $^{30)}$
which is too small to explain the LSND result.\\

The standard scenario does not solve the atmospheric neutrino problem.
One can sacrifice the HDM suggesting that some other particles
are responsible for the structure formation in the Universe,
or consider strongly degenerate neutrino spectrum which has a potential
problem with neutrinoless double beta decay. In both cases no appreciable
effects in KARMEN/LSND experiments  are expected.\\

\noindent
{\it 2. Neutrinos and light singlet fermion}

More safed way to accommodate all the anomalies is to introduce
one new neutrino state$^{31 - 37)}$.
As follows from LEP bound on the
number of neutrino species this state
 should be sterile (singlet of standard group).
Taking  into account also
strong bound on  parameters of oscillations into sterile neutrino from
Primordial Nucleosynthesis one
can write the following ``scenario" $^{31 - 35)}$:\\

\noindent
(i) Sterile neutrino has the mass $m_S \sim (2 - 3) \cdot 10^{-3}$ eV
and mixes with $\nu_e$, so that the resonance conversion
$\nu_e -\nu_s$ solves the solar neutrino problem;\\
(ii)The masses of  ${\nu}_{\mu}$ and  ${\nu}_{\tau}$
are in the range 2 - 3 eV,  they supply
the desired hot component of the DM; \\
(iii)  ${\nu}_{\mu}$ and  ${\nu}_{\tau}$
form the pseudo Dirac neutrino with large (maximal) mixing and
the oscillations ${\nu}_{\mu} - {\nu}_{\tau}$
explain the atmospheric neutrino problem;\\
(iv) $\nu_e$ is very light:  $m_1 < 2\cdot10^{-3}$ eV. The
$\bar{\nu}_{\mu} - \bar{\nu}_e$ mixing can be strong enough
to explain the LSND result.\\
(v) Production of heavy elements in
supernova via  ``r-processes" is problematic for this scenario.\\

What is the origin of sterile neutrino?
Of course, the RH neutrino
components are natural candidates.
However in this case  the see-saw mechanism does not operate.
Another possibility is that  singlet fermion $S$
exists beyond the standard see-saw structure$^{36)}$.
Its appearance  is motivated by some reasons not related to the
neutrino physics.
And moreover,  this scalar can be  family blind.
The lightness of $S$ is nontrivial since
standard model symmetry does not protect
$S$  from  acquiring the mass $m_s \gg m_W$.\\

We suggest$^{36)}$ (see also$^{37)}$)
that $S$  mixes with active neutrinos via
interactions with RH neutrino components only. So that in the
basis $(S, \nu_e, N)$ the mass matrix has the following form
\begin{equation}
{\cal M} = \left( \matrix{
0 & 0 & m_{es}\cr
0 & 0 & m_e \cr
m_{es} & m_e & M_e
\cr}\right).
\end{equation}
Diagonalization  gives
one  massless  and one light state
with mass
$
m_1 \simeq -(m_e^2 + m_{es}^2)/ M_e\;.
$
The $\nu_e$--$S$ mixing angle is determined by
$ \tan \theta_{es} = m_e / m_{es}$.
This mechanism allows one to generate  simultaneously the mass and
mixing without introduction of very small mass scale.
Taking for $m_e$ the typical Dirac mass of the first generation:
$m_e \sim (1-5)$ MeV, and suggesting that $\nu_e \rightarrow S$
conversion explains the solar neutrino problem, we find
$m_{es}
= {m_e \over \tan\theta_{es}} \simeq (0.02-0.3)$ GeV$^{36)}$.

How the  scale $0.1 - 1$ GeV appears in singlet sector?
One possibility  is the  supersymmetry endowed by some spontaneously
broken global symmetry, e.g. $U(1)_G$ in the simplest case. Spontaneous
violation of $U(1)_G$ results in goldstone boson, and in
the supersymmetric limit  corresponding superpartner (fermion)
is  massless. Supersymmetry breaking
parametrized by soft breaking
terms leads in general to the $S$-mass which can be
as big as
$O(m_{3/2})$. That is the supersymmetry alone can not protect very small
mass scales, and one needs some additional care to further suppress
$m_S$.

One  possibility is the based on $R$-symmetry ($G \equiv R$)
spontaneously broken up to the R-parity$^{36)}$.
The  R-parity conservation   requires for the fermion $S$ to be
a component of singlet superfield which has no VEV.
This allows one to construct a simple model  in which the
properties (mass and mixing) of $S$ follow from the conservation of
R-symmetry.  $S$ is mixed with RH neutrinos by the
interaction with  additional singlet field $y$ which can acquire VEV
radiatively after soft SUSY breaking.  The model can naturally
incorporate the spontaneous violation of Peccei-Quinn symmetry or/and lepton
number.  The fields involved can spontaneously generate the $\mu$--term.
Approximate horizontal (family) $U(1)^h$ symmetry
can provide simultaneous explanations for the predominant coupling of $S$ to
the first generation (thus satisfying the Nucleosynthesis bound) and for the
pseudo-Dirac
structure of $\nu_{\mu}$--$\nu_{\tau}$ needed in solving the atmospheric
neutrino and hot dark matter problem.
Breaking of $U(1)^h$ can be arranged in such a way that the
parameters of $\bar{\nu}_\mu \to \bar{\nu}_e$ oscillations are in the
region of sensitivity of  KARMEN and LSND experiments.\\

\noindent
{\large Conclusion}\\

After top quark discovery the neutrinos are the only known
fermions with unknown
masses. There is some hope that things are comming to a head
and in 2 - 3 years we will know the  answer.\\

\centerline{REFERENCES}
\vspace*{0.2cm}
\baselineskip=.3truecm
\noindent
1. K. Lande, Nucl.\ Phys.\ B38 (Proc. Suppl.) (1995) 47.\\
2. K. Inoe, these Proceedings.\\
3. S. Eliott, these Proceedings.\\
4. C. Tao, these Proceedings.\\
5. I. Barabanov,  (1988) unpublished\\
6. J. N. Bahcall and H. A. Bethe, Phys. Rev. Lett., 65 (1990) 2233;
Phys. Rev. D47 (1993)
\phantom{dftg}1298.\\
7. M. Spiro  and D. Vignaud,  Phys. Lett. B 242 (1990) 279.\\
8. V. Castellani, S. Degl'Innocenti and G. Fiorentini, Phys. Lett.
B 303 (1993) 68: Astron.
\phantom{dftg}Astroph. 271 (1993) 601.\\
9. H. Hata, S. Bludman and P. Langacker,  Phys. Rev. D 49 (1994) 3622.\\
10. V. S. Berezinsky, Comm Nucl. Part. Phys. 21 (1994) 249.\\
11. A. Yu. Smirnov, Report DOE/ER/40561-136 -INT94-13-01.\\
12. W. Kwong and S. P. Rosen, Phys. Rev. Lett., 73 (1994) 369.\\
13. S. Parke, Phys. Rev. Lett. 74 (1995) 839.\\
14. S. Degl'Innocenti, G. Fiorentini and M. Lissia,  Ferrara preprint
INFNFE-10-94.\\
15. V. S. Berezinsky, these Proceedings.\\
16. N. Hata and P. Langacker, Phys. Rev. D 50 (1994) 632;
Report UPR -0625T, hep-
\phantom{dftg}ph/9409372.\\
17. S. P. Krastev and A. Yu. Smirnov Phys. Lett. B 338 (1994) 282.\\
18. V. S. Berezinsky G. Fiorentini and M. Lissia, Phys. Lett., B 341
(1994) 38.\\
19. J. N. Bahcall and M. H. Pinsonneault, Rev. Mod. Phys. 60 (1989) 297.\\
20. A. Dar and G. Shaviv, Report Technion-Ph-94-5 (1994).\\
21. Z. G. Berezhiani and A. Rossi, Phys. Rev. D51 (1995) 5229.\\
22. P. I. Krastev and S. P. Petcov, report SISSA-9/95/EP.\\
23. A. Yu. Smirnov, D. N. Spergel and J. N. Bahcall,
Phys. Rev. D49 (1994) 1389.\\
24. C. R. Proffitt, Astrophys. J. 425 (1994) 849;
A. Kovetz, G. Shaviv,  Astrophys. J. 426
\phantom{dftg}(1994) 787;
J. N. Bahcall and M. H. Pinsonneault, IASSNS-AST 95/24.\\
25. M. Fukugita and T. Yanagida, Phys. Lett., B174 (1986) 45.\\
26. F. Vissani and A. Yu. Smirnov, Phys. Lett., B 341 (1994) 173.\\
27. A. Brignole, H. Murayama and R. Rattazzi, Phys. Lett. B335 (1994) 345.\\
28. Y. -Z. Qian et al., Phys. Rev. Lett., 71 (1993) 1965.\\
29. S. Dimopoulos, L. J.  Hall and S Raby, Phys. Rev.  D47 (1993)
R3697.\\
30. K. S. Babu, J.C. Pati and F. Wilczek, Preprint IASSNS-HEP 95/37.\\
31. D.\ O.\ Caldwell and R.\ N.\ Mohapatra, Phys.\ Rev.\ D48
(1993) 3259.\\
32. J.\ Peltoniemi and J.\ W.\ F.\ Valle, Nucl.\ Phys.\ B406
(1993) 409.\\
33. J.\ Peltoniemi, D.\ Tommasini, and J.\ W.\ F.\ Valle,
Phys.\  Lett.\ B298 (1993) 383;\\
34. J. T. Peltoniemi,  Mod.~Phys.~Lett. A8 (1993) 3593.\\
35. J.\ R.\ Primack, J.\ Holtzman, A.\ Klypin, and D.\ O.\ Caldwell,
Phys.~Rev.~Lett.\ 74 (1995)
\phantom{dftg}2160; and references therein.\\
36. E. J. Chun, A. S. Joshipura and A. Yu. Smirnov,
IC/95/76, PRL-TH/95-7, hep-ph/9505275.\\
37. E. Ma and P. Roy, preprint UCRHEP-T145 (hep-ph/9504342).\\


\end{document}